\documentclass[preprint,authoryear]{elsarticle}
\usepackage{amssymb,amsmath,epsfig,epstopdf}
\usepackage{paralist}
\usepackage{setspace}
\usepackage{lineno}
\usepackage{graphicx}

\def\av#1{\left\langle#1\right\rangle}
\def\to{\rightarrow}
\def\Var{{\rm Var}\,}

\def\A{\textbf{A}}
\def\B{\textbf{B}}

\begin{document}


\title{A note on the distribution of admixture segment lengths and ancestry proportions under pulse and two-wave admixture models}

\author[huji]{Shai Carmi\corref{cor}}
\ead{shai.carmi@huji.ac.il}
\author[cu]{James Xue}
\author[cu]{Itsik Pe'er}

\address[huji]{Braun School of Public Health, The Hebrew University, Jerusalem, Israel}
\address[cu]{Department of Computer Science, Columbia University, New York, NY, USA}
\cortext[cor]{Corresponding author}

\begin{abstract}

Admixed populations are formed by the merging of two or more ancestral populations, and the ancestry of each locus in an admixed genome derives from either source. Consider a simple ``pulse'' admixture model, where populations $\A$ and $\B$ merged $t$ generations ago without subsequent gene flow. We derive the distribution of the proportion of an admixed chromosome that has $\A$ (or $\B$) ancestry, as a function of the chromosome length $L$, $t$, and the initial contribution of the $\A$ source, $m$. We demonstrate that these results can be used for inference of the admixture parameters. For more complex admixture models, we derive an expression in Laplace space for the distribution of ancestry proportions that depends on having the distribution of the lengths of segments of each ancestry. We obtain explicit results for the special case of a ``two-wave'' admixture model, where population $\A$ contributed additional migrants in one of the generations between the present and the initial admixture event. Specifically, we derive formulas for the distribution of $\A$ and $\B$ segment lengths and numerical results for the distribution of ancestry proportions. We show that for recent admixture, data generated under a two-wave model can hardly be distinguished from that generated under a pulse model.

\end{abstract}

\maketitle

\section{Introduction}

Present-day genomes are mosaics of ancestries from the different sources that merged to form modern populations (e.g., \cite{HapMix,PCAdmix,RFMix}). Estimating the time of each admixture event and the relative contribution of each source population is an important problem in population genetics. To estimate admixture times, \cite{Johnson_ssPCA} fitted the number of ancestry switches; \cite{Pugach_2011}, and later \cite{CoxWavelet2015}, matched simulations to the typical segment length, as estimated from a wavelet transform of the local ancestry along the genome; and \cite{MigrationIBD_Nielsen}, as well as \cite{Gravel}, fitted the distribution of segment lengths. However, these methods require an accurate identification of the boundaries of admixture segments, which is not always available, in particular for computationally phased data. Reich and colleagues \citep{ROLLOFF,Patterson_AdmixTools,ALDER,PickrellAfricaAdmixture} fitted the decay of admixture linkage disequilibrium (LD) with genetic distance, but such methods can be confounded by background LD. \cite{Myers_Science2014} recently proposed a promising approach based on the probability of two fixed loci to have given ancestries. Gene flow parameters can also be inferred using more general demographic inference methods, e.g., based on the allele frequency spectrum \citep{Excoffier_2013,dadi} or segment sharing \citep{Palamara_migrations}; however, to use these methods one must specify and infer a model for the entire history.

Recently, Rosenberg and colleagues \citep{Verdu,Goldberg2014,Goldberg2015}, \cite{Liang2014}, and \cite{Gravel} derived analytical results for the moments of the ancestry proportion, namely, the fraction of a chromosome that descend from a given source population. These ancestry proportions can be reliably inferred (e.g., \cite{ADMIXTURE,Structure}), and the derived moments have been used for admixture time inference (e.g., \cite{HennNorthAfrica2,Liang008078,LiangPacific}). However, these methods did not make use of the entire distribution, as no analytical results were available. Here, we derive first results for the distribution of the ancestry proportions. We obtain an explicit formula for the case of a ``pulse'' admixture event, and we demonstrate its application to the inference of the admixture parameters. We then derive an expression in Laplace space for a general admixture model of arbitrary complexity, but which requires knowledge of the distribution of admixture segment lengths. For the special case of a ``two-wave'' admixture, we obtained the segment length distribution, and consequently, through numerical Laplace inversion, the distribution of ancestry proportions.

\section{The distribution of ancestry proportions under pulse admixture}

Consider the pulse model, where an admixed population formed $t$ generations ago as a result of merging of populations $\A$ and $\B$, and where the proportions of ancestry contributed by $\A$ and $\B$ were $m$ and $1-m$, respectively. Under this model, each locus in a chromosome of a present-day admixed individual can trace its origin to $\A$ or $\B$ with probabilities $m$ and $1-m$, respectively. We assume lineages break apart along the chromosome, due to recombination, at rate $t$ per Morgan. Ignoring genetic drift and the underlying pedigree, we assume that upon recombination, the new source population is selected at random. Therefore, a recombination event will lead to a change of ancestry from $\A$ to $\B$ with probability $1-m$ and from $\B$ to $\A$ with probability $m$. The lengths chromosomal segments with $\A$ ancestry will thus be exponentially distributed with rate $(1-m)t$, and similarly for the $\B$ segments (rate $mt$) \citep{Gravel}. We neglect the first generation after admixture, where $\A$ and $\B$ segments do not yet mix \citep{Gravel}. As pointed out by \cite{Liang2014}, the assumption of independent and exponentially distributed segment lengths breaks down for very short and very long times since admixture, due to the effect of the underlying pedigree and the accumulation of genetic drift, respectively. Nevertheless, for the relevant time-scales and effective population sizes of many human populations (around 10-200 generations and effective size in the thousands), segment lengths should be extremely well approximated by independent exponentials.

Given a chromosome of length $L$ (Morgans), the ancestry along the chromosome can be modeled as a two-state process with states $\A$ and $\B$, and with the distribution of segment lengths in each state given above. We are interested in the distribution of $x$, the fraction of the chromosome in state $\A$. Adopting a result of \cite{Stam1980}, the desired distribution is given by
\begin{align}
\label{pulse_pdf}
f(x;L)& =(1-m)e^{-mh}\delta(x) + me^{-(1-m)h}\delta(1-x) \\ &+ 
m(1-m)he^{-h\left[(1-m)x+m(1-x)\right]} \times \nonumber \\ & \qquad \qquad \times 
\left\{\frac{mx+(1-m)(1-x)}{\alpha}I_1(2h\alpha)+2I_0 (2h\alpha)\right\} \nonumber,
\end{align}              
where $h\equiv tL$, $\alpha\equiv \sqrt{m(1-m)x(1-x)}$, and $I_0$ and $I_1$ are the modified Bessel functions of the first kind of order 0 and 1, respectively. Note the delta functions at $x=0$ and $x=1$, corresponding to the probability of the entire chromosome having $\B$ only or $\A$ only ancestry, respectively. The mean ancestry proportion satisfies $\av{x}=m$, as expected. The variance is given by
\begin{equation}
\Var[x]=\frac{2m(1-m)}{h^2}\left(e^{-h}+h-1\right), 
\end{equation}
in agreement with Eq. (A16) in \cite{Gravel}. Note that Eq. \eqref{pulse_pdf} represents the distribution of the ancestry proportion across repetitions of the ancestral process (for a single chromosome), rather than across chromosomes from a sample \citep{Gravel,Liang008078}. However, unless the population is very small (compared to the admixture time and the sample size), histories of different chromosomes are to a good approximation independent, and the two distributions are the same.

\section{Inference of admixture times}

\begin{figure}
\vspace{-3.5cm}
\includegraphics[scale=0.4]{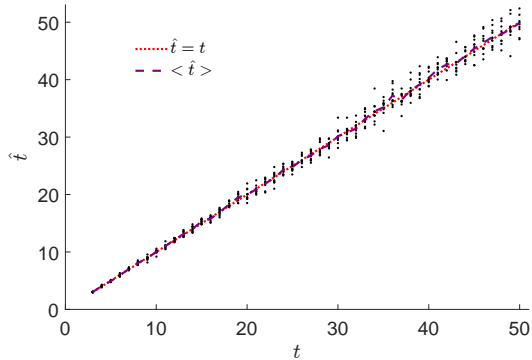}
\vspace{-3.5cm}
\caption{Inference of admixture times using the distribution of ancestry proportions. We simulated an admixture pulse history under the Markovian Wright-Fisher model of \cite{Gravel}. The model assumes that the $2N$ haploid chromosomes in the current generation are formed by following a Markovian path within the $2N$ chromosomes of the previous generation. Ancestry changes occur as a Poisson process with rate 1 (per Morgan). Each chromosome in the first generation is assigned to population $\A$ or $\B$ with probabilities $m$ and $1-m$, respectively, and the evolution of the chromosomes is traced for $t$ generations. We used $m=0.5$, $L=2$M, and $N=2500$, and varied $t$. Ancestry proportions from pairs of chromosomes were averaged to generate diploid individuals. We then set the inferred $m$ to the mean $\A$ ancestry, and used the distribution of ancestry proportions, Eq. \eqref{pulse_pdf}, to infer the admixture time $t$. Each dot in the plot shows the inferred time, $\hat{t}$, for one simulation. The dotted red line corresponds to $\hat{t}=t$, and the dashed purple line to the mean inferred time, $\av{\hat{t}}$.}
\label{fig_pulse}
\end{figure}

In theory, given the observed ancestry proportion for each chromosome in a sample, Eq. \eqref{pulse_pdf} can be used to compute the likelihood of the observed data for given admixture parameters. In practice, in the absence of trios or pedigree information, phase switches are abundant, and hence, it is difficult to accurately determine the ancestry proportion per chromosome. However, it is still possible to determine the diploid ancestry proportion, $y=(x_1+x_2)/2$. Given that homologous chromosomes have independent histories, the diploid ancestry proportion distribution, $f_d(y;L)$, can be computed from Eq. \eqref{pulse_pdf} by convolution. Suppose we are now given the diploid ancestry proportions $y_{ij}$ for individuals $i=1,...,n$ and for chromosomes $j=1,...,22$ (where each chromosome has length $L_j$). Assuming chromosomes are independent both within and between individuals, the likelihood of the data is given by
\begin{equation}
\label{diploid_like}
\textrm{likelihood}=\prod_{i=1}^n\prod_{j=1}^{22} f(y_{ij};L_j)
\end{equation}
Maximum likelihood estimates (MLE) for $m$ and $t$ can then be obtained by a simple grid search. Simulation results with perfect knowledge of segment boundaries demonstrated that the method can infer correctly both $m$ and $t$ (Figure \ref{fig_pulse}), although the variance increases with $t$, as expected. Coalescent simulations followed by inference of ancestry proportions using \textsc{ADMIXTURE} \citep{ADMIXTURE} and application of our method demonstrated again high accuracy, at least as long as the $\A$ and $\B$ population were sufficiently diverged (not shown). However, when $\A$ and $\B$ are closely related, the distributions of the true and inferred ancestry proportions may differ, affecting the accuracy of the method.

\section{The distribution of ancestry proportions under a general distribution of segment lengths}

We have so far considered a simple pulse model, under which the distributions of segment lengths are exponential with known rates. Under a more complex admixture history, we assume that the distributions of the lengths of $\A$ and $\B$ segments take the general form $q_{\A}(\ell)$ and $q_{\B}(\ell)$. We still assume that $\A$ and $\B$ segments are independent (see below). The process can then be modeled as a two-state process. We start on the left end of the chromosome in state $\A$ or $\B$ with probabilities $p_A=\av{\ell_A}/\left(\av{\ell_A}+\av{\ell_B}\right)$ and $1-p_A$, respectively (where $\av{\ell_A}$ and $\av{\ell_B}$ are the mean segment lengths), and draw a random segment length from the selected ancestry. When the first segment terminates, we switch ancestries and draw a segment length from the other ancestry, and so on until we reach the end of the chromosome.

The distribution of $x$, the $\A$ ancestry proportion, can be computed in Laplace space by extending renewal theory methods developed in the physics domain (e.g., \cite{GodrecheLuck,Margolin2004}). Let $s$ be the Laplace pair of $L$ (the total chromosome length) and $u$ as the Laplace pair of $L_A\equiv xL$ (the total chromosome length covered by $A$ segments). We transform the density $f(L_A;L)$ (from which the density of $x$ can be easily obtained) to $\hat{f}(u;s)$, and after some calculations, we obtain
\begin{equation}
\label{phi_s_u}
\hat{f}(u;s)=\frac{s\left[1-\hat{q}_{\A}(s+u)\hat{q}_{\B}(s)\right]+u\left[1-\hat{q}_{\B}(s)\right]\left\{ 1-p_A\left[1-\hat{q}_{\A}(s+u)\right]\right\} }{s(s+u)\left[1-\hat{q}_{\A}(s+u)\hat{q}_{\B}(s)\right]}.
\end{equation}
In the above equation, $\hat{q}_{\A}(s)$ and $\hat{q}_{\B}(s)$ are the Laplace transforms ($\ell\to s$) of $q_{\A}(\ell)$ and $q_{\B}(\ell)$, respectively. The details of the derivation are somewhat tedious and therefore omitted. It can be shown, using Eq. \eqref{phi_s_u}, that the mean ancestry proportion $\av{x}$ approaches $p_A$ as $L \to \infty$. It can be also shown that Eq. \eqref{phi_s_u} reduces to Eq. \eqref{pulse_pdf} for the admixture pulse model.

\section{Conditions under which consecutive segments are independent}

To obtain concrete results for complex admixture histories, we use the model developed by \cite{Gravel} (section \emph{General incoming migration in the absence of drift} and Figure 3 there). Gravel proposed that the ancestry along the chromosome could be described by a Markov process, whose states correspond to the identity of the source population (i.e., $\A$ or $\B$), combined with the time when each segment entered the admixed population. Gravel then derived the transition rates for any admixture history. While the extended state space process is Markovian under any history, consecutive $\A$ and $\B$ segment lengths are generally no longer independent. However, little thought reveals that as long as migration beyond the the initial event is limited to one population, consecutive segment lengths remain independent.

\section{The distribution of segment lengths under a two-wave admixture model}

Consider a model where populations $\A$ and $\B$ have merged $t_1$ generations ago, contributing proportions $m$ and $1-m$ to the admixed population. Then, $t_2$ ($<t_1$) generations ago, migrants from population $\A$ have replaced a proportion $\mu$ of the gene pool of the admixed population. No other events then take place until the present. The corresponding Markov process, using the method of \cite{Gravel}, has three states: $A_1$, $A_2$, and $B$, representing migrant segments from $\A$ at time $t_1$, from $\A$ at time $t_2$, and from $\B$ (at time $t_1$), respectively. Let us compute the distributions of the lengths of $\A$ and $\B$ segments.

The transition rate is $t_1$ when at states $A_1$ and $B$, and $t_2$ when at $A_2$. It can be shown that once a transition is made, the next state is chosen according to the following transition probability matrix
\begin{equation}
\label{eq_trans_mat}
\mathbf{P}=\left(
\begin{matrix}
m\left(1-\mu\frac{t_2}{t_1}\right) & \mu\frac{t_2}{t_1} & (1-m)\left(1-\mu\frac{t_2}{t_1}\right) \\ 
m(1-\mu) & \mu & (1-m)(1-\mu) \\ 
m\left(1-\mu\frac{t_2}{t_1}\right) & \mu\frac{t_2}{t_1} & (1-m)\left(1-\mu\frac{t_2}{t_1}\right)
\end{matrix}
\right).
\end{equation}
The states are ordered as $(A_1,A_2,B)$ and $\mathbf{P}_{ij}$ ($i,j=1,2,3$) is the probability to jump from state $i$ to state $j$. Note again that we neglected the first generation after admixture, during which $\A$ and $\B$ segments do not yet mix \citep{Gravel}.

It is now easy to see that $\B$ segment lengths are distributed exponentially with rate $t_1(1-\mathbf{P}_{B,B})$, or
\begin{equation}
\label{eq_qB_final}
q_{\B}(\ell)=t_1\left[1-(1-m)\left(1-\mu\frac{t_2}{t_1}\right)\right]\exp \left\{-t_1\ell\left[1-(1-m)\left(1-\mu\frac{t_2}{t_1}\right)\right]\right\}.
\end{equation}
This equation was also (implicitly) derived by \cite{Ni2015} in a different way. For the $\A$ segments, define $q_{A_1}(\ell)$ as the distribution of $\A$ segment lengths, \emph{when the process entered the $\A$ states at state $A_1$}, and similarly for $q_{A_2}(\ell)$. Since the process always enters $A_1$ and $A_2$ from $B$ (ignoring the leftmost end of the chromosome), the distribution of $\A$ segments therefore satisfies
\begin{equation}
\label{eq_PAx_general}
q_{\A}(\ell)=\frac{\mathbf{P}_{B,A_1}}{1-\mathbf{P}_{B,B}}q_{A_1}(\ell)+\frac{\mathbf{P}_{B,A_2}}{1-\mathbf{P}_{B,B}}q_{A_2}(x).
\end{equation} 
To find $q_{A_1}(\ell)$ and $q_{A_2}(\ell)$, we write integral equations,
\begin{align}
& q_{A_1}(\ell)=\mathbf{P}_{A_1,B}t_1e^{-t_1 \ell}+\int_0^{\ell}t_1e^{-t_1y}\left[\mathbf{P}_{A_1,A_1}q_{A_1}(\ell-y)dy+\mathbf{P}_{A_1,A_2}q_{A_2}(\ell-y)\right]dy \nonumber \\
& q_{A_2}(x)=\mathbf{P}_{A_2,B}t_2e^{-t_2x}+\int_0^{\ell}t_2e^{-t_2y}\left[\mathbf{P}_{A_2,A_1}q_{A_1}(\ell-y)dy+\mathbf{P}_{A_2,A_2}q_{A_2}(\ell-y)\right]dy.
\end{align}
We solve these equations by using a Laplace transform ($\ell \to s$) and the convolution theorem,
\begin{align}
& \hat{q}_{A_1}(s)=\frac{t_1}{t_1+s}\left[\mathbf{P}_{A_1,B}+\mathbf{P}_{A_1,A_1}\hat{q}_{A_1}(s)+\mathbf{P}_{A_1,A_2}\hat{q}_{A_2}(s)\right] \nonumber \\
& \hat{q}_{A_2}(s)=\frac{t_2}{t_2+s}\left[\mathbf{P}_{A_2,B}+\mathbf{P}_{A_2,A_1}\hat{q}_{A_1}(s)+\mathbf{P}_{A_2,A_2}\hat{q}_{A_2}(s)\right].
\end{align}
These are two linear equations in two variables ($\hat{q}_{A_1}(s)$ and $\hat{q}_{A_2}(s)$), which are easily solved. Then, $q_{A_1}(\ell)$ and $q_{A_2}(\ell)$ are obtained by Laplace transform inversion. We then use Eq. \eqref{eq_PAx_general} to obtain $q_{\A}(\ell)$. We carried out these steps in \textsc{Mathematica}, leading to the result
\begin{align}
\label{eq_qA_final}
q_{\A}(\ell)=\frac{(1-m)e^{-\gamma \ell/2}\left[C_1\sinh(\beta\ell/2)+C_2\cosh(\beta \ell/2)\right]}{\beta\left[mt_1+\mu t_2(1-m)\right]},
\end{align}
where $\gamma=t_1+(1-m)(t_1-t_2\mu)$, $\beta=\sqrt{\gamma^2-4t_1t_2(1-m)(1-\mu)}$,
\begin{equation*}
C_1=m^2(t_1-\mu t_2)^3-m(t_1-\mu t_2)\left[t_1^2-t_1 t_2-2t_2^2\mu(1-\mu)\right]+t_2^2\mu(1-\mu)\left[t_1-t_2(1-\mu)\right],
\end{equation*}
and
\begin{equation*}
C_2=\left[m(t_1-\mu t2)^2+\mu(1-\mu)t_2^2\right]\beta.
\end{equation*}
We note that we can also view the second migration wave as gene flow coming from a \emph{third} population. Our results then automatically provide the distribution of ancestry proportions coming from each of the three sources.

\begin{figure}
\vspace{-5cm}
\includegraphics[scale=0.5]{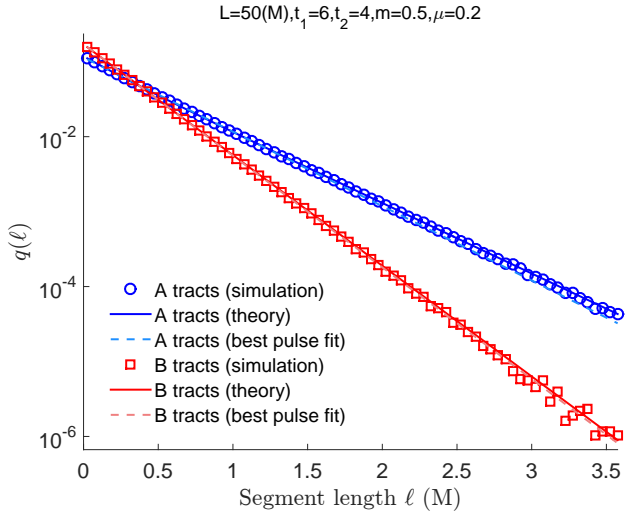}
\vspace{-3.5cm}
\caption{Two-wave admixture: simulations and theory for the segment length distribution. We simulated a two-wave admixture model according to a Markovian Wright-Fisher model (\cite{Gravel}; as in Figure \ref{fig_pulse}) with $N=2500$. The other model parameters are indicated on top of the figure. We used a particularly long chromosome to avoid boundary effects. We recorded the lengths of segments that descend from $\A$ and $\B$ populations, and plotted their histogram (circles and squares, respectively). The theoretical distributions, $q_{\A}(\ell)$ and $q_{\B}(\ell)$ (Eqs. \eqref{eq_qA_final} and \eqref{eq_qB_final}, respectively), are plotted as solid lines. We then fitted a pulse admixture model with just two parameters ($m$ and $t$) by matching the means of the empirical $\A$ and $\B$ segment lengths. The distributions of $\A$ and $\B$ segment lengths under the pulse model (exponentials with rates $(1-m)t$ and $mt$, respectively) are plotted as dashed light-colored lines. The best fit for $t$ was 5.7, intermediate between $t_1$ and $t_2$.}
\label{fig_two_wave_tracts}
\end{figure}

We ran simulations of the two-wave model under the Markovian Wright-Fisher framework described by \cite{Gravel} (see Figure \ref{fig_pulse}). Representative simulation results are shown in Figure \ref{fig_two_wave_tracts}. It can be seen that our theory matches the empirical data very well. We note, though, that the empirical distributions can be fitted quite well to an admixture pulse model with parameters $m_{\textrm{pulse}}$ equal to the expected mean ($\mu+m(1-\mu)$) and $t_{\textrm{pulse}}$ intermediate between $t_1$ and $t_2$. This suggests that, at least for admixture parameters tested here, any inference based on the more complex model may not have sufficient evidence to justify the additional gene flow event (see also \cite{Myers_Science2014}).

\section{The distribution of ancestry proportions for two-wave admixture}

\begin{figure}
\vspace{-1cm}
\includegraphics[scale=0.7]{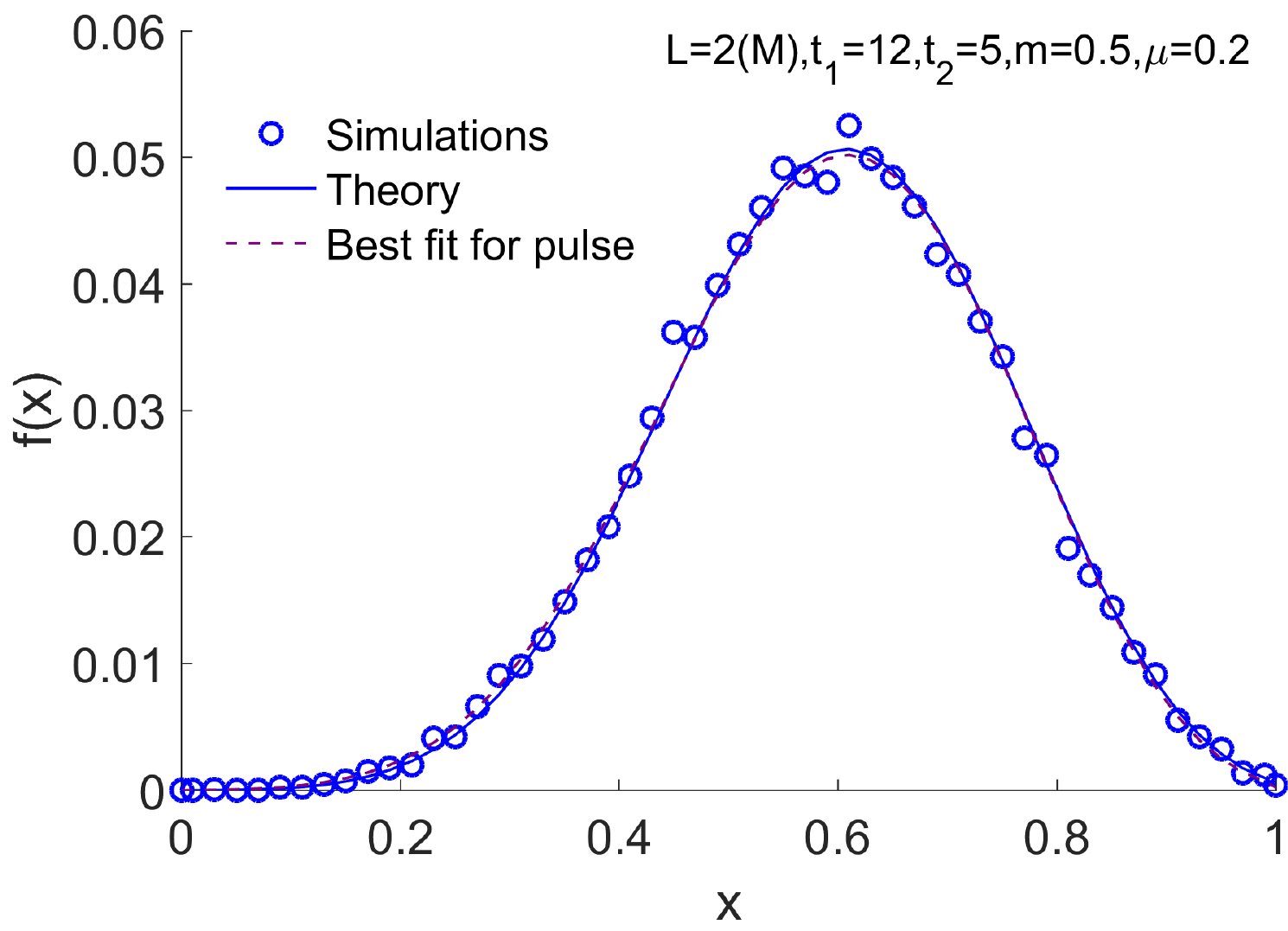}
\caption{Two-wave admixture: simulations and theory for the ancestry proportions. We simulated a two-wave admixture model according to a Markovian Wright-Fisher model (\citep{Gravel}; as in Figure \ref{fig_pulse} but without averaging pairs of haplotypes) with $N=2500$. The other model parameters are indicated on top of the figure. We recorded the fraction of each chromosome that descends from the $\A$ population, and plotted the histogram (circles). The theory (based on Eq. \eqref{phi_s_u}) is plotted as a solid (blue) line. We then fitted a pulse admixture model with just two parameters ($m$ and $t$) by matching the mean and variance of the empirical data. The distribution of the ancestry proportions under the pulse model (Eq. \eqref{pulse_pdf}) is plotted as a dashed (purple) line. The best fit for $t$ was 9.7, intermediate between $t_1$ and $t_2$.}
\label{fig_two_wave}
\end{figure}

Now that we have $q_{\A}(\ell)$ and $q_{\B}(\ell)$ for the two-wave model (Eqs. \eqref{eq_qA_final} and \eqref{eq_qB_final}, respectively), we can use Eq. \eqref{phi_s_u} for the distribution of the ancestry proportions. We inverted $\hat{f}(u;s)$ with respect to $u$ using \textsc{Mathematica} and then numerically with respect to $s$, to obtain $f(x;L)$. Simulation results are shown in Figure \ref{fig_two_wave}, demonstrating that our theoretical results fit the empirical data very well. Here too, excellent fit is achieved also by the pulse admixture model.

\section{Discussion}

We proposed a simple method for inference of admixture parameters based on the empirical distribution of the proportion of each chromosome that descend from each source population. One advantage of this approach is not having to rely on the precise boundaries of the admixture segments. Rather, we only need the total amount of genetic material from each source, which is typically easier to estimate, even without explicitly performing local ancestry inference (e.g., using programs such as \textsc{ADMIXTURE}). Additionally, our method is easily adapted to unphased data, and enjoys the advantages of maximum likelihood estimation. Finally, we were able to completely generalize the results to the case of two-wave admixture, where gene flow from one of the populations occurred in two different occasions. 

Extending the results to additional gene flow events (limited to a single source population) should be straightforward, at least numerically. However, extensions to more complex models or to continuous migration seem complicated. The finite population size could generally be neglected, as long as it is much larger than the sample size and (numerically) than the number of generations since admixture, which is typically the case. For small populations, genetic drift has two contrasting effects. The first is to increase the variance of the distribution of the ancestry proportions, since the potential for a ``back-coalescence'' implies that recombination events do not always change the ancestry, effectively increasing segment lengths. Based on the analysis of the SMC' model by \cite{Liang2014}, Eq. \eqref{fig_pulse} should still hold, but with $h=tL$ replaced by $h=\tau L$, where $\tau=2N\left(1-e^{-t/(2N)}\right)$ (derivation not shown; note that $\tau \to t$ for $t\ll 2N$). The second effect arises when the distribution is over a sample from the population, reducing the variance due to the fact that lineages may coalesce already before reaching the time of admixture. A complete theory is yet to be developed, perhaps along the lines of \cite{Liang008078}.

While our results provide reasonable accuracy for a parameter regime typical of some natural populations, we caution that often, the method may not be directly applicable. Intuitively, the information exploited by our method is mostly in the first moments of the distribution; while we estimate parameters in a more principled MLE approach, our method is still prone to inaccuracies in estimating ancestry proportions, as in studies based merely on the variance (e.g., \citep{HennNorthAfrica2} and our unpublished results for Ashkenazi Jews). Nevertheless, our method may serve as a building block (or a sanity check) for more complex approaches. For example, our results already clearly demonstrate the inherent infeasibility of distinguishing a pulse and a two-wave admixture histories using segment lengths and ancestry proportions statistics for recent admixture (past $\approx 10$ generations). Finally, our theoretical results will be of great interest to researchers in population genetics and coalescent theory working in the very active field of admixture modeling.

\section*{Software}

\textsc{Matlab} code is available for the inference of $m$ and $t$ for the pulse admixture model, as well as for the distributions of segment lengths and ancestry proportions for the two-wave model. See {https://github.com/scarmi/admixture\textunderscore models}.

\section*{Acknowledgements}

S.C. thanks the Human Frontier Science Program for financial support.

\section*{References}

\bibliographystyle{elsarticle-harv}
\bibliography{../Admixture}

\end{document}